\documentclass[fleqn,usenatbib]{mnras}

\usepackage{newtxtext,newtxmath}

\usepackage[T1]{fontenc}
\usepackage{ae,aecompl}

\usepackage{graphicx}	
\usepackage{amsmath}	
\usepackage{amssymb}	

\title[X-ray Analysis of the A3667 SE Relic]{XMM-Newton Observations of the Southeastern Radio Relic in Abell 3667}

\author[Storm, Vink, Zandanel, Akamatsu]{
Emma Storm,$^{1}$\thanks{E-mail: e.m.storm@uva.nl}
Jacco Vink,$^{1,2,3}$
Fabio Zandanel,$^{1}$
Hiroki Akamatsu$^{3}$
\\
$^{1}$GRAPPA, University of Amsterdam, Science Park 904, 1098 XH Amsterdam, Netherlands\\
$^{2}$API, University of Amsterdam, Science Park 904, 1098 XH Amsterdam, Netherlands\\
$^{3}$SRON Netherlands Institute for Space Research, Utrecht, The Netherlands
}

\date{Accepted XXX. Received YYY; in original form ZZZ}

\pubyear{2017}

\begin{document}
\label{firstpage}
\pagerange{\pageref{firstpage}--\pageref{lastpage}}
\maketitle

\begin{abstract}

Radio relics, elongated, non-thermal, structures located at the edges of galaxy clusters, are the result of synchrotron radiation from cosmic-ray electrons accelerated by merger-driven shocks at the cluster outskirts. However, X-ray observations of such shocks in some clusters suggest that they are too weak to efficiently accelerate electrons via diffusive shock acceleration to energies required to produce the observed radio power. We examine this issue in the merging galaxy cluster Abell 3667 (A3667), which hosts a pair of radio relics. While the Northwest relic in A3667 has been well studied in the radio and X-ray by multiple instruments, the Southeast relic region has only been observed so far by \textit{Suzaku}, which detected a temperature jump across the relic, suggesting the presence of a weak shock. We present observations of the Southeastern region of A3667 with \textit{XMM-Newton} centered on the radio relic. We confirm the existence of an X-ray shock with Mach number of about 1.8 from a clear detection of temperature jump and a tentative detection of a density jump, consistent with previous measurements by \textit{Suzaku}. We discuss the implications of this measurement for diffusive shock acceleration as the main mechanism for explaining the origin of radio relics. We then speculate on the plausibility of alternative scenarios, including re-acceleration and variations in the Mach number along shock fronts.
\end{abstract}
%



\section{Introduction}

The growth of structure at large scales is driven by violent, energetic mergers of galaxy clusters. These events result in turbulence and shocks that propagate throughout the cluster volume. Observations of diffuse synchrotron emission in clusters indicate that some fraction of the energy generated by merger events is transferred into non-thermal components, including magnetic fields and cosmic rays, in the intracluster medium (ICM) \citep[for recent reviews, see e.g.,][]{2012A&ARv..20...54F,2014IJMPD..2330007B}. X-ray observations reveal that the hot thermal plasma that makes up the ICM in merging clusters is generally highly disturbed, as indicated by surface brightness discontinuities and complicated temperature profiles \citep[e.g.,][]{2007PhR...443....1M}.

Diffuse radio emission from clusters is typically classified by morphology \citep{2012A&ARv..20...54F}. Radio halos are characterized by $\sim$Mpc-scale, unpolarized emission that tends to be centered on the center of the cluster, and are thought to be the result of turbulence throughout the ICM. Radio relics are found near the outskirts of merging clusters, often in pairs on opposite sides of the cluster, and tend to be highly elongated and moderately polarized at the $15-30\%$ level. Radio relics are some of the best evidence for the existence of merger shocks in clusters \citep[e.g.,][]{1998A&A...332..395E,2012SSRv..166..187B}.

If radio relics are the result of shock-driven particle acceleration, then these shocks should also be identifiable in X-ray observations of the ICM via a surface brightness discontinuity and corresponding temperature jump at the location of the shock. However, identification of these shocks is hampered by the fact that they are often in the outskirts of clusters, where the thermal gas density is low and the X-ray emission is therefore faint. 

A fruitful approach to identify merger shocks so far has been to perform deep X-ray observations of clusters with radio relics. While there are currently over 50 observed radio relics, only a handful of those have an associated a temperature jump or surface brightness discontinuity observed in the X-ray band that suggests a shock front \citep[recently, e.g.,][]{2013PASJ...65...16A,2016MNRAS.461.1302E,2016arXiv160607433S}. In contrast, there are also several clusters with detected X-ray shocks that are not associated with radio relics (although they do host other diffuse radio emission such as halos; e.g. the western shock of the Bullet Cluster: \cite{2002ApJ...567L..27M,2014MNRAS.440.2901S}; A520: \cite{2005ApJ...627..733M,2014A&A...561A..52V}; A2146: \cite{2010MNRAS.406.1721R,2017arXiv170803641H}). Additionally, several shocks were found in merging clusters via surface brightness and temperature jumps in the \textit{Chandra} archive by \citet{2017arXiv170707038B}; however, none of these appear to be associated with radio relics (and only some host radio halos).

Abell 3667 (hereafter A3667; $z=0.0556$\footnote{Redshift from the NASA Extragalactic Database (NED).}) is a well-known merging cluster in the southern hemisphere that hosts a pair of radio relics to the Northwest (NW) and Southeast (SE) \citep{1997MNRAS.290..577R,2003PhDT.........3J,2013MNRAS.430.1414C,2014MNRAS.445..330H,2015MNRAS.447.1895R}. X-ray observations of the central region of the cluster indicate that it is highly disturbed with a prominent cold front \citep{2001ApJ...551..160V,2004A&A...426....1B,2014ApJ...793...80D,2017MNRAS.467.3662I}. The NW relic is one of the most powerful observed radio relics \citep{2012A&ARv..20...54F} and the encompassing region has been extensively studied in the X-ray. \textit{XMM-Newton} (hereafter, \textit{XMM}) and \textit{Suzaku} observations indicate the existence of a shock front at the location of the NW relic \citep{2010ApJ...715.1143F,2012PASJ...64...49A,2016arXiv160607433S}. However, the SE relic region has only been observed in the X-ray band so far with \textit{Suzaku} \citep{2013PASJ...65...16A}. A temperature jump was measured with \textit{Suzaku} at the location of the SE relic, suggesting the presence of a weak shock with a Mach number of $1.75\pm 0.13$ \citep{2013PASJ...65...16A}. However, due to the broad PSF of Suzaku, point sources were not excluded from that analysis, which could lead to over- or underestimates of the cluster temperature profile across the relic region. Additionally, discontinuities in the surface brightness are critical to confirm the existence and specific location of a shock front. The resolution of Suzaku is too poor to have measured a such a discontinuity, as it would smear out any discontinuities on arcminute scales.

The non-thermal radio emission from some shocks implies that these shocks are accelerating particles, probably by the mechanism of diffusive shock acceleration (DSA; \citealt{2001RPPh...64..429M}). However, the lack of radio relics associated with some shocks suggest that for DSA to be efficient, certain conditions have to be fulfilled, probably regarding the shock Mach number and strength and orientation of the local magnetic field. It is generally assumed that for DSA to be efficient the Mach number has be $M\gtrsim 3$ \citep[e.g.][]{2011ApJ...734...18K}. In fact, \citet{2014ApJ...780..125V} showed that there exists a minimum shock strength required for particle acceleration via DSA of $\mathcal{M}=\sqrt{5}\approx2.2$, which is larger than many Mach number estimates for relic shocks \citep[e.g.,][]{2014MNRAS.440.3416O,2016ApJ...818..204V,2016MNRAS.461.1302E,2016MNRAS.460L..84B}, including the recent Mach number estimate of the SE relic in A3667 \citep{2013PASJ...65...16A}. 

We present here an analysis of a new observation of the SE relic region in A3667 with \textit{XMM}, to provide independent confirmation of the existence and strength of this X-ray shock first discovered by \citet{2013PASJ...65...16A} and to test the viability of DSA as the prevailing mechanism for accelerating electrons at the shocks associated with radio relics. With this new observation, we can confirm the low Mach number estimate from Suzaku via a temperature jump and exploit the high spatial resolution of XMM to identify a corresponding surface brightness jump and pinpoint the location of the shock front, which is not typically possible from a spectral analysis of the temperature jump alone. We show in Figure~\ref{fig:XrayMosaic} a mosaic of all publicly-available \textit{XMM} observations of A3667, including this new SE observation, overlaid with the radio relics as observed with the Sydney University Mongolo Sky Survey (SUMSS) \citep{1999AJ....117.1578B,2003MNRAS.342.1117M}.

This paper is structured as follows. In Section~\ref{sec:obs}, we outline the observation and data reduction strategy. In Sections~\ref{sec:spec} and \ref{sec:spat}, we describe the spectral and spatial analysis procedures, respectively. In Section~\ref{sec:dis}, we discuss the results of spectral and spatial fitting. Finally, we conclude in Section~\ref{sec:con}. We use a cosmology with the following parameter values: $\Omega_{\text{M}}=0.3, \Omega_{\Lambda}=0.7, H_0=70$~km~s$^{-1}$~Mpc~$^{-1}$. For these values, at a redshift of $0.0556$, $1'' = 1.08$~kpc. Unless otherwise noted, uncertainties are reported at the $68\%$ confidence level.

\section{Observation and Data Reduction}\label{sec:obs}

\begin{figure}
  \centering
  \includegraphics[clip,scale=0.25]{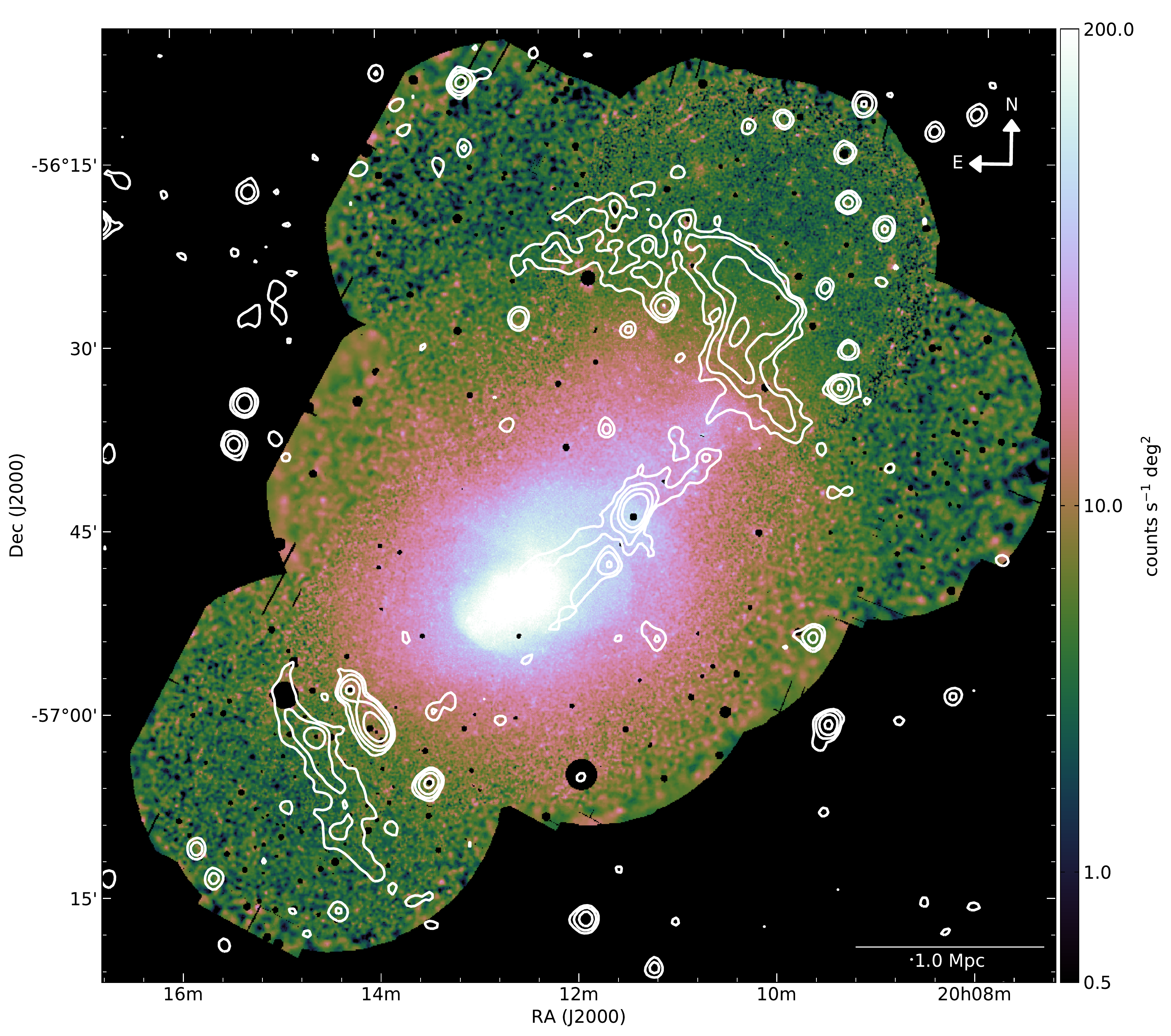}
  \caption{Background-subtracted, exposure-corrected, adaptively-smoothed \textit{XMM} image of A3667 in the $0.5-4$~keV range. This image is a mosaic of all publicly-available observations of A3667 in the \textit{XMM} archive. The new observation that covers the SE relic is in the lower left corner. Overlaid in white are the radio relic contours from SUMSS at $843$~MHz. Contour levels are $[2,6,18,54]$~mJy/beam and are smoothed to $\sim1$~arcmin for clarity. The compass in the upper right corner of the figure indicates the orientation of this and all following images in the paper.}
  \label{fig:XrayMosaic}
\end{figure}

We  present here the analysis of a new, 76ks \textit{XMM} EPIC observation of the SE region of A3667 that is roughly centered on the radio relic (ObsID: 0761210101; PI: Vink). We used the Extended Source Analysis Software (ESAS), part of the \textit{XMM} Science Analysis System (SAS; version 15.0.0), to reduce the observation. Following the suggested analysis procedure in the ESAS Cookbook\footnote{\url{heasarc.gsfc.nasa.gov/docs/xmm/esas/cookbook/xmm-esas.html}}, we first created clean event files for the MOS1, MOS2, and pn detectors, and filter the events for flares. We used the ESAS task \textit{cheese} to remove point sources from the observation. The total clean exposure is 61ks each for the MOS1 and MOS2 cameras, and 45ks for the pn camera. We then generated quiescent particle background (QPB) images and spectra, which are modeled from closed-filter-wheel data and normalized using the unexposed corners of our observation in the ESAS framework. We show a background-subtracted, exposure-corrected image of the observation in Figure~\ref{fig:SE_SBprofile_wedge}. All of the observations for the mosaic in Figure~\ref{fig:XrayMosaic} were analyzed with ESAS in a similar fashion.

\section{Spatial Analysis}\label{sec:spat}
\subsection{Filtered Images}\label{sec:spat_1}
\begin{figure*}
  \centering
  \includegraphics[clip,scale=0.25]{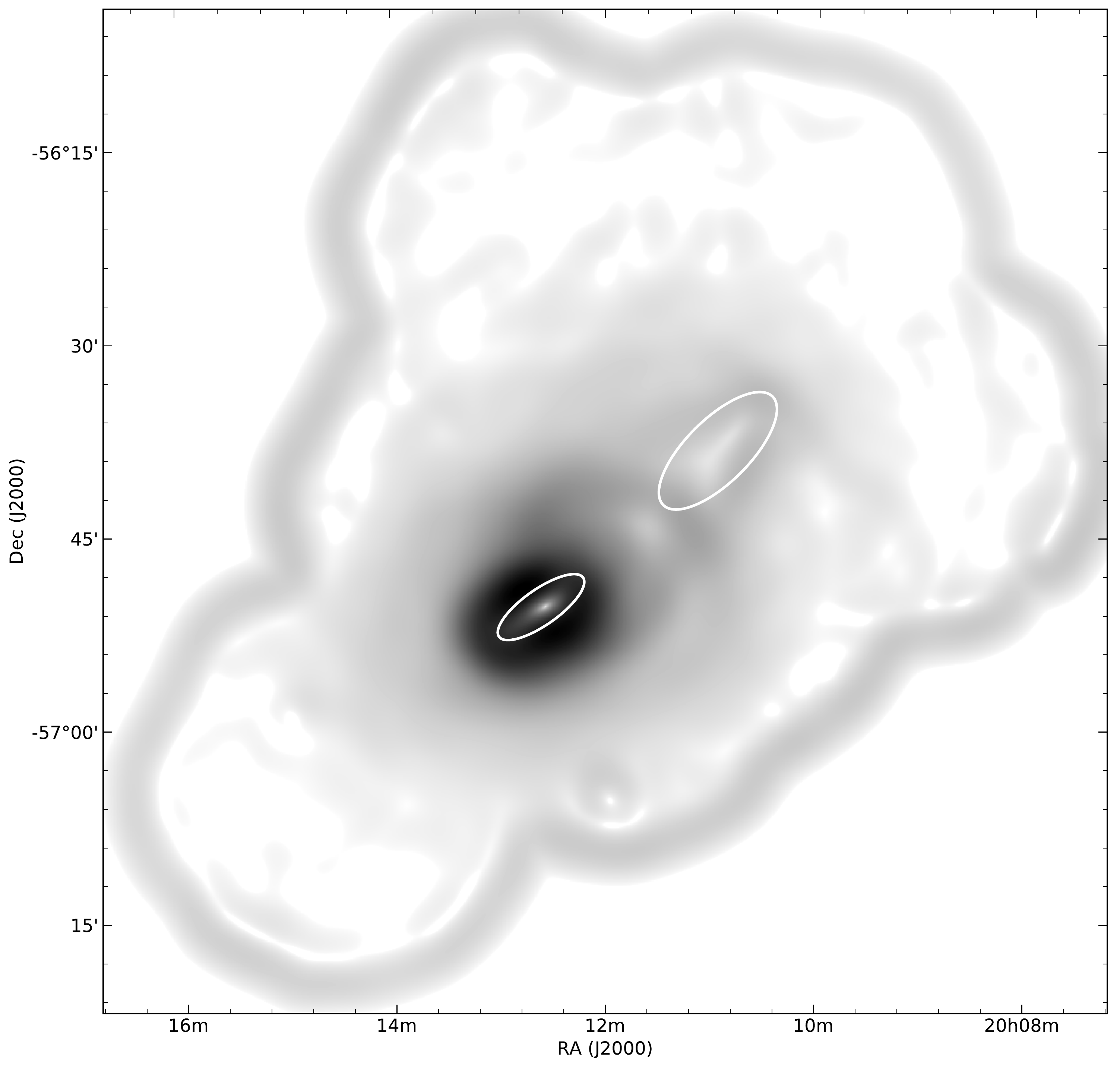}
     \includegraphics[clip,scale=0.25]{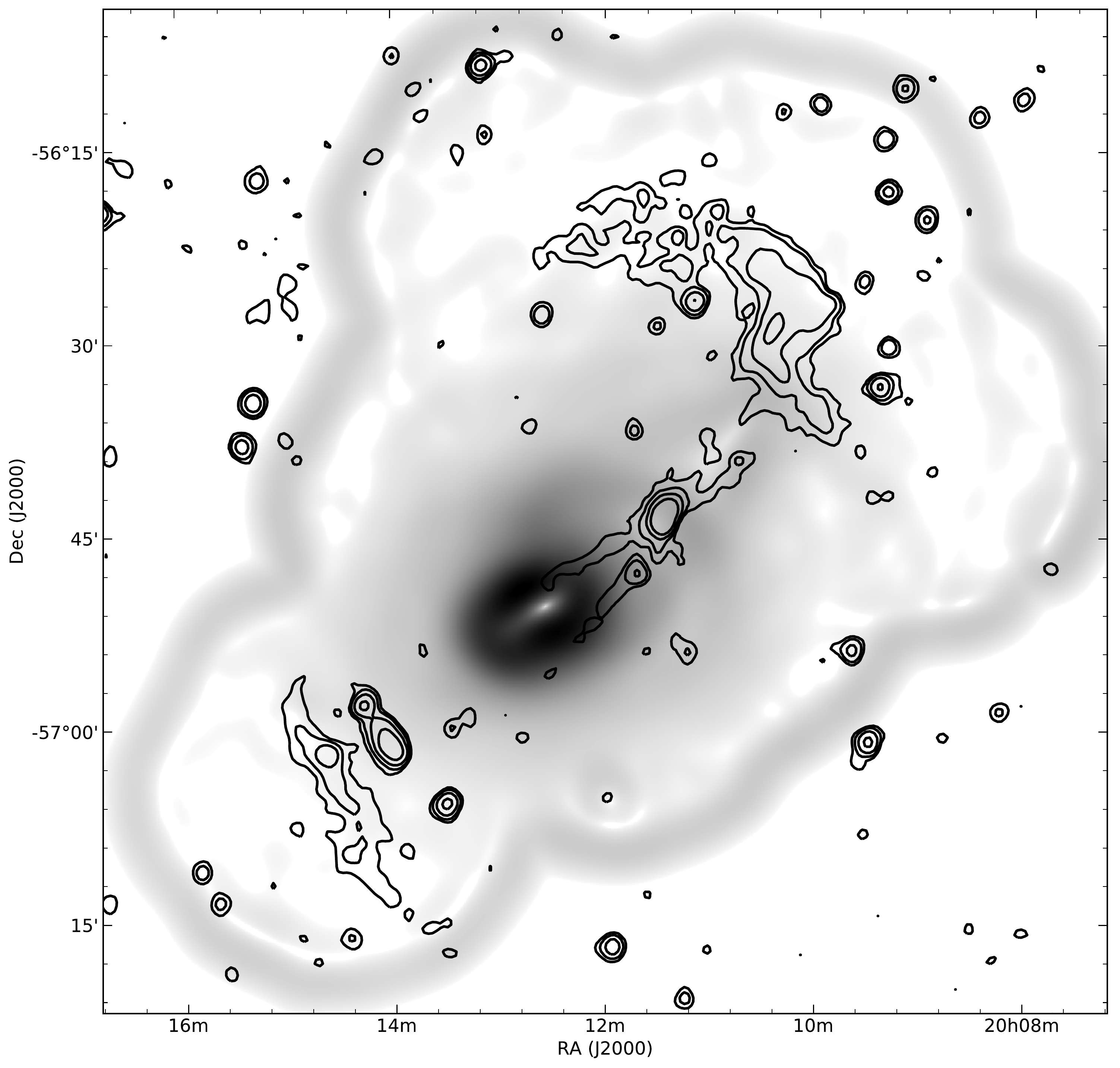}
  \caption{Left: GGM filtered image, $\sigma=16$, $0.5-4$~keV.  The two white ellipses mark the locations of the ``plateau" (center of image) and the ``mushroom" (towards the northwest) features described in Section~\ref{sec:spat_1}. Right: Same as left, but with SUMSS contours overlaid.}
  \label{fig:SE_filters}
\end{figure*}

A surface brightness discontinuity is one of the strongest indicators of a shock front, coupled with a spectral analysis. However, no discontinuity is obvious from a visual inspection of the SE relic region in Figure~\ref{fig:XrayMosaic}. In order to better highlight any discontinuity, we apply an image processing technique designed to enhance edge-like features called the Gaussian Gradient Method (GGM). When applied to X-ray images, the GGM filter can effectively find sharp discontinuities at different scales, including shocks, by varying the Gaussian width of the filter \citep{2016MNRAS.460.1898S,2016MNRAS.461..684W}. We tried several widths, and found that a width of $\sigma=16$ works best for highlighting edges in the low surface brightness regions near the relics. We show this filtered image in Figure~\ref{fig:SE_filters}. Edges near both the NW and SE relics can be clearly seen. 

We also recover the same features found by \citet{2016MNRAS.460.1898S} when analyzing a \textit{Chandra} image of A3667, including the so-called ``inner plateau" (this is observed more clearly for smaller widths, which we do not show). The so-called ``mushroom" pointed out by \citet{2016arXiv160607433S} towards the NW is also seen in this filtered image.

\subsection{Surface Brightness Profile}

\begin{figure}
  \centering
  \includegraphics[clip,scale=0.25]{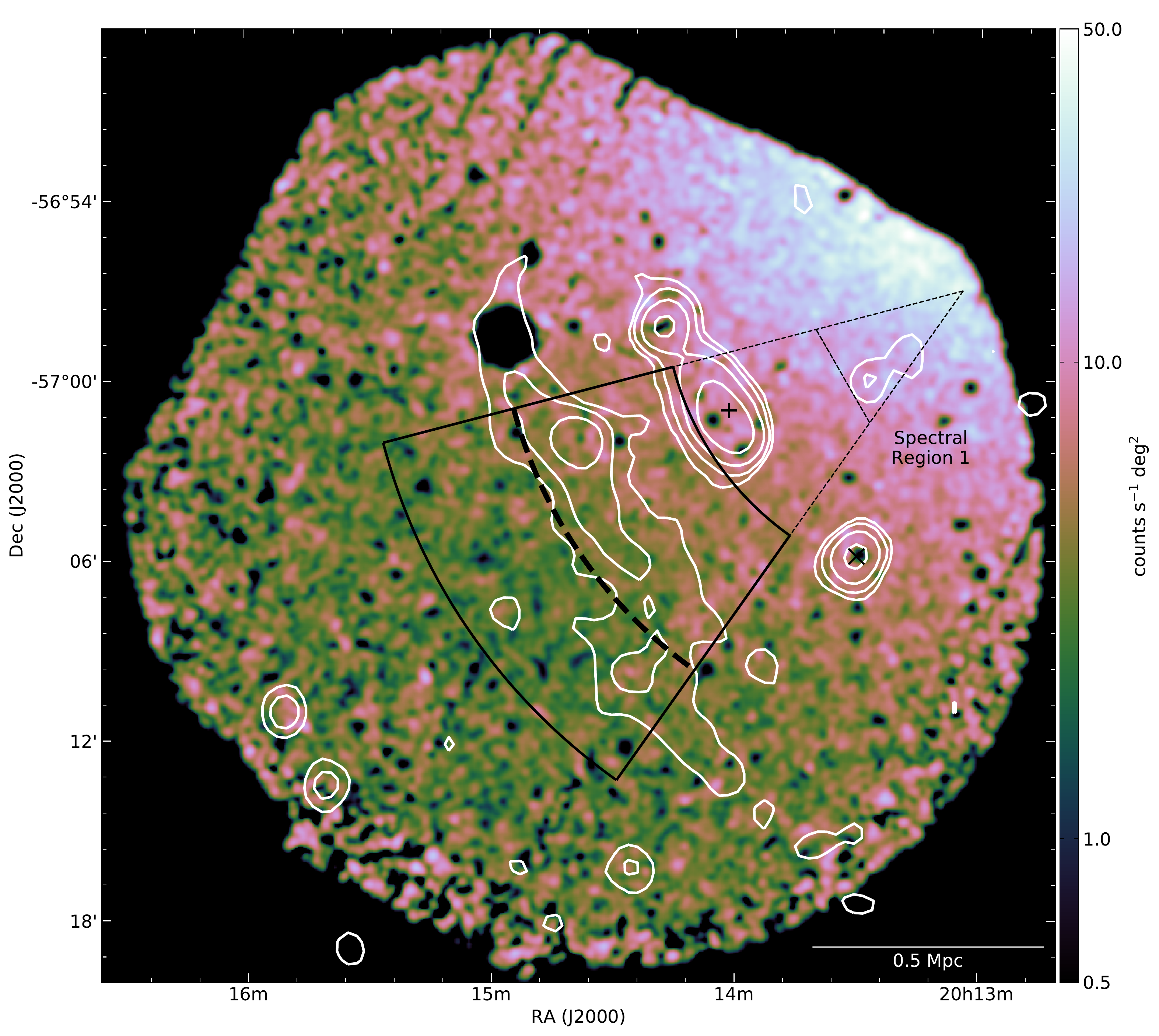}
  \caption{\textit{XMM} image of the SE observation of A3667, $0.5-4$~keV, smoothed with a Gaussian kernel with a FWHM of $10$~arcsec. The white contours show the southern relic with the same contour levels as in Figure~\ref{fig:XrayMosaic}. The solid black sector shows the region used to extract the surface brightness profile. The dashed black line indicates the best-fit position for the shock radius. The dashed wedge towards the northwest of the image is shown to indicate the origin in Figure~\ref{fig:SBProfile}; that is, the point of the wedge corresponds to $r=0$. The coordinates of this point are ($20^h13^m04.32^s$, $-56^d57^m0^s$). The straight dashed black line cutting through the wedge labeled "Spectral Region 1" marks the location of the beginning of the spectral regions in Figure~\ref{fig:SE_spectra_regions} and $r=0$ on Figure~\ref{fig:TempProfile}. The black ``+'' shows the position of the point source PMN J2014-5701 and the black ``x'' marks the position of SUMSS J201330-570552, discussed in Section~\ref{sec:dis2}.}
  \label{fig:SE_SBprofile_wedge}
\end{figure}

While the GGM-filtered image does suggest that there is a surface brightness discontinuity in the region of the SE relic, we would like to make quantitative statements about its strength and position. To do this, we extract a surface brightness profile in the region marked by the sector in Figure~\ref{fig:SE_SBprofile_wedge}. For this analysis, we consider the energy range $0.5-4$~keV, which is highly unlikely to be contaminated by residual soft proton emission (however, the results do not significantly change if the $0.5-7$~keV or $0.5-2$~keV bands are used instead).

We model the thermal electron density with a broken power law:

\begin{align}
\label{eq:SB}
\begin{split}
n_2(r) &= \mathcal{C} n_0 \left(\frac{r}{r_{\text{shock}}} \right)^{-\alpha_2},\quad r\leq r_{\text{shock}}\\
n_1(r) &= n_0 \left(\frac{r}{r_{\text{shock}}} \right)^{-\alpha_1},\quad r>r_{\text{shock}}\,;
\end{split}
\end{align}
where $n_0$ is the electron number density normalization, $C \equiv n_2/n_1$ is the shock compression ratio , $r_{\text{shock}}$ is the shock radius, and $\alpha_1$ and $\alpha_2$ are the power law indices of the post- and pre-shock regions, respectively.

We use the Proffit package\footnote{\url{http://www.iasf-milano.inaf.it/~eckert/newsite/Proffit.html}} \citep{2011A&A...526A..79E} for fitting. Proffit uses the technique in Appendix A of \citet{2009ApJ...704.1349O} to numerically project the density profile above along the line of sight to the surface brightness. The functional form of the actual model that is fit with Proffit is given by:

\begin{align}
\begin{split}
S(r_{\perp}) &= \mathrm{norm} \int n(r)^2 \mathrm{d}\ell + \mathrm{const}\,;
\end{split}
\end{align}
where $S(r_{\perp})$ is the surface brightness along the projected 2D radius $r_{\perp}$, norm is the normalization, $n(r)$ is the 3D density profile given in Equation~\ref{eq:SB}, and $\ell$ is the line-of-sight such that $r^2 = r_{\perp}^2 + \ell^2$. There are a total of six model parameters: the normalization and an optional constant background term (both in units of surface brightness, counts~s$^{-1}$~arcmin$^{-2}$) and four additional parameters from the density profile: the shock compression ratio $C = n_2/n_1$, the shock jump radius $r_\mathrm{shock}$, and the two power law indices $\alpha_1$ and $\alpha_2$. The data are grouped into bins with a minimum signal-to-noise of 5 and are logarithmically spaced, varying in size from about 10-30~arcsec across the region considered.
\begin{figure}
  \centering
  \includegraphics[clip,scale=0.7]{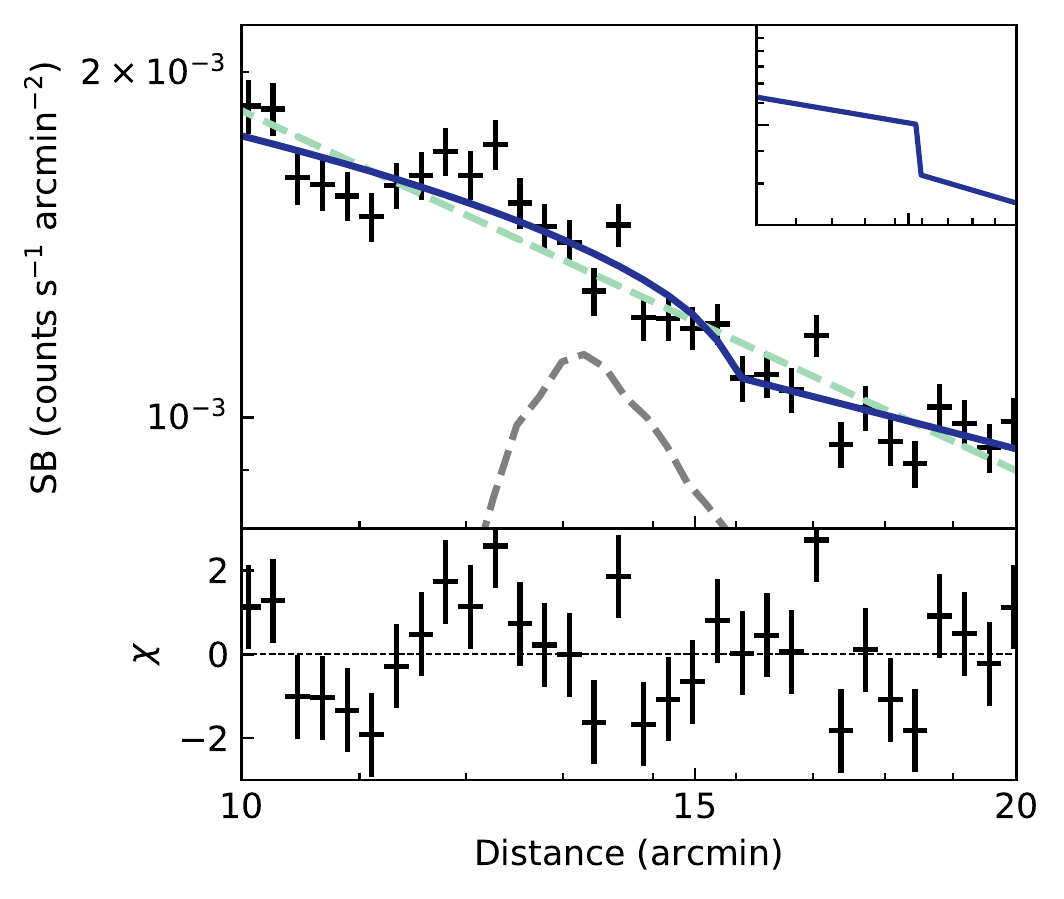}
  \caption{Surface brightness profile extracted from the sector in Figure~\ref{fig:SE_SBprofile_wedge}. The best-fit broken-power-law model is shown in blue. The best-fit single-power-law is represented by the light green dashed line. The gray dashed line is the SE radio relic surface brightness profile, extracted from the same sector as the X-ray emission, and is shown in arbitrary units of surface brightness (the peak is approximately 10 times the lowest shown value). A radius of $10$~arcmin corresponds to the narrower edge of the solid black sector shown in Figure~\ref{fig:SE_SBprofile_wedge}. The 3D density model as a function of 3D distance is shown in the inset.}
  \label{fig:SBProfile}
\end{figure}
The surface brightness profile and best-fit model is shown in Figure~\ref{fig:SBProfile}. The signal-to-background ratio is relatively low, especially in the region outside of the relic, so the overall fit is somewhat poor with a reduced $\chi^2/dof=51/27 \approx 1.9$.  The best-fit value for the compression ratio is: $\mathcal{C} = 1.4\pm0.2$. The best-fit shock radius is at the outer edge of the radio relic, $r_{\text{shock}}=15.5\pm0.2$~arcmin in Figure~\ref{fig:SBProfile}. This corresponds to $r \sim 10.5$~arcmin in Figure~\ref{fig:TempProfile}. However, a broken power law does yield the best fit compared to a single power law ($\chi^2/dof=67/29$; this is a power-law fit directly to the surface brightness and not to the density). An F-test reveals that the improvement of the broken power law fit to the density over a single power law fit to the surface brightness is at the 2 sigma level, i.e.
it has an associated chance probability of $2.5\%$ (the F-statistic is 4.2).

The sector was chosen to optimize both the size of the jump and overall quality of the fit. Varying the bin size or minimum signal-to-noise did not significantly change the quality of the fit or parameter values. The length of the sector did have an impact on the fit. Outside the sector shown in Figure~\ref{fig:SE_SBprofile_wedge}, the surface brightness profile flattens out towards the SE. This is likely due to cluster emission becoming fainter than the sky background, as was also seen in the spectral fitting of the outer regions. The outer edge of the sector was chosen by eye to be where this transition appears to occur. The inner edge of the sector does not strongly affect the best-fit values of the fit, but does affect the overall quality of the fit. When a constant background term is added to Equation~\ref{eq:SB} and left free, the formally best-fit value is negative, which is unphysical. If this constant term is forced to be zero or positive, the best-fit value is consistent with zero. The $2\sigma$ upper limit on a constant background term is $<1.5\times10^{-4}$ counts~s$^{-1}$~arcmin$^{-2}$. This value is consistent with the count rate of the sky background estimated from the spectral fit in Section~\ref{sec:spec}.

We can relate the Mach number, $\mathcal{M}$, and compression ratio, $C$, assuming $\gamma=5/3$, using the Rankine-Hugoniot jump conditions:

\begin{equation}
\mathcal{M} = \left(\frac{3C}{4-C}\right)^{1/2}\,.
\end{equation}

For our best-fit value of $C=1.4\pm0.2$, this yields $\mathcal{M}=1.3\pm0.1$. The uncertainties reported are purely statistical and do not take into account systematics due to sector choice.

\begin{table}
\centering
\caption{Background model components}\label{tab:BkgSpec}
\begin{tabular}{cccc}
	\hline
	Component & $\Gamma$ & $kT$ & Normalization \\
    \hline
    LHB & -- & $0.184 \pm 0.001$ & $(1.84 \pm 0.02) \times 10^{-6}$ \\
    GH  & -- & $0.63 \pm 0.02 $ & $(3.5 \pm 0.1)\times 10^{-7}$ \\
    CXB & $1.41$ & -- & $(5.6 \pm 0.2) \times 10^{-7}$ \\
    RESP & $0.76 \pm 0.03$ & -- & $(1.9 \pm 0.1) \times 10^{-3}$ \\
	\hline
\end{tabular}
\begin{flushleft}
X-ray temperatures have units of keV. The normalizations are in XSPEC units: $\frac{10^{-14}}{4\pi (D_A(1+z)^2)}\int n_e n_H dV$, where $D_A$ is the angular diameter distance to the source, $z$ is the redshift, and $n_e$ and $n_H$ are the electron and hydrogen densities (cm$^{-3}$). The spectral index for the CXB was kept fixed. The RESP normalization listed is for the pn detector, region 7. The other normalizations are scaled to this value by the flux in a given detector and region reported by the ESAS command \textit{proton\_scale}, and vary in the range $(0.5-2) \times 10^{-3}$.
\end{flushleft}
\end{table}

\section{Spectral Analysis}\label{sec:spec}
\subsection{Model Description}

\begin{figure}
  \centering
  \includegraphics[clip,scale=0.25]{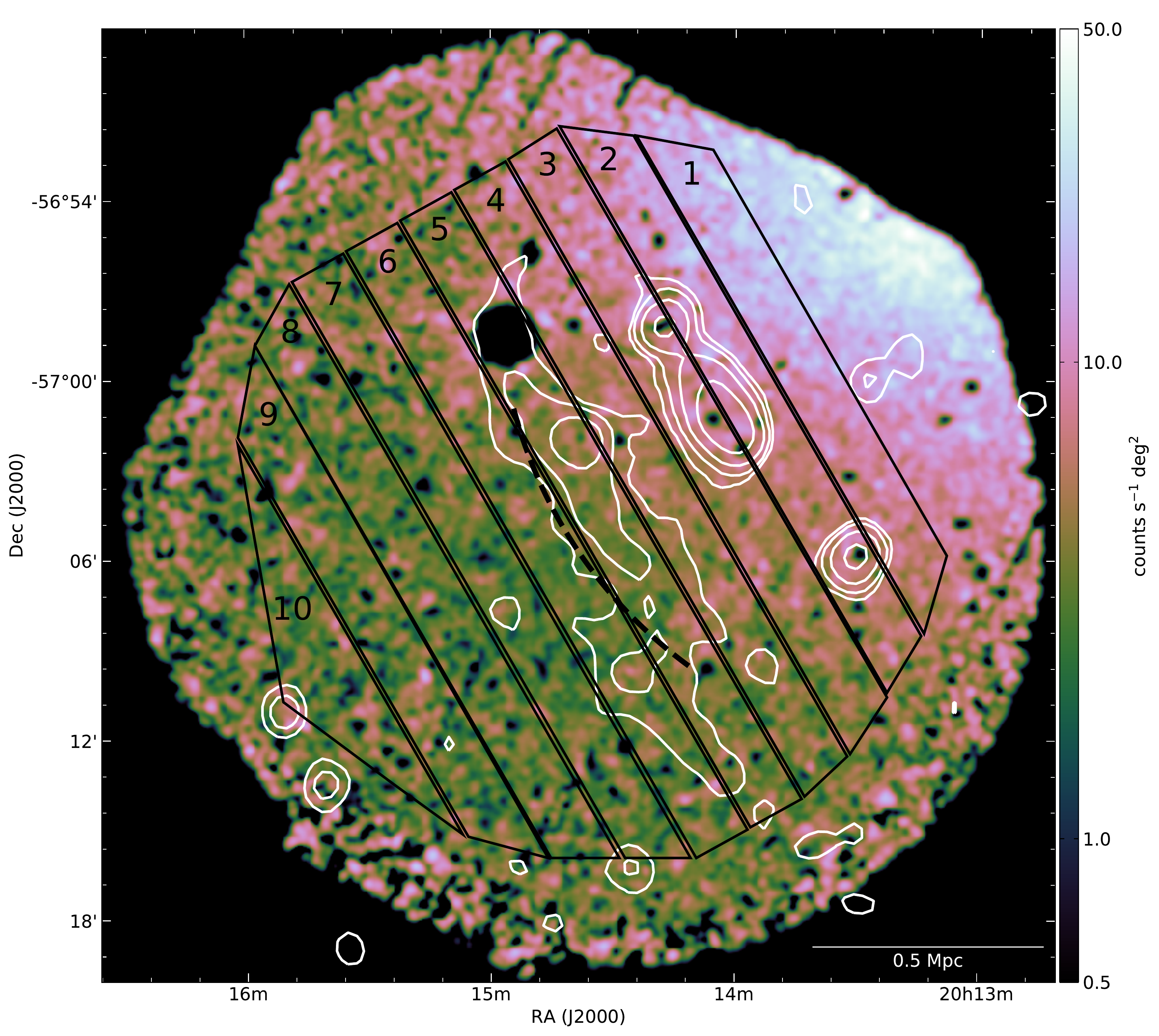}
  \caption{\textit{XMM} image of the SE region of A3667, $0.5-4$~keV, smoothed with a Gaussian kernel with a FWHM of $10$~arcsec. Overlaid in black are the regions used in spectral fitting. The dotted black line marks the location of the best-fit shock radius from the spatial analysis. The white contours show the southern relic (with the same contour levels as in Figure~\ref{fig:XrayMosaic}).}
  \label{fig:SE_spectra_regions}
\end{figure}

We perform a spectral analysis in the regions shown in  Figure~\ref{fig:SE_spectra_regions} to search for a temperature jump across the SE radio relic. These 10 regions are $2$~arcmin wide, and were chosen to follow the shape of the relic and their placement was informed by the location of the density break detected from the spatial analysis. The spectra are extracted from the regions in Figure~\ref{fig:SE_spectra_regions} and binned to 50 counts per bin. We use XSPEC (version 12.9.0) for spectral fitting with $\chi^2$ statistics. For the MOS1 and MOS2 detectors, we fit the spectra over the energy range $0.5-10$~keV. Due to several instrumental lines in the pn detectors above $7$~keV, we restrict the spectral fit for the pn detector to $0.5-7$~keV.

We build our spectral model as the sum of background (instrumental and astrophysical) and target emission components.  The astrophysical emission component,
i.e. the emission from the cluster itself, is modeled with an APEC thin plasma model \citep{2001ApJ...556L..91S}. We fix the redshift to $0.0556$ and the abundance to $0.2$ solar\footnote{using the abundance table from \citet{1989GeCoA..53..197A}} across all regions and detectors, which is typical for cluster outskirts \citep{2008A&A...487..461L}. The normalization and cluster temperature are left free in each region, but are linked across detectors.

For the astrophysical background, we include the unabsorbed Local Hot Bubble (LHB) and the absorbed Galactic Halo as APEC\footnote{\url{www.atomdb.org}} thin plasma models \citep{2001ApJ...556L..91S}, and the absorbed Cosmic X-ray background (CXB) as a power law in our fit. In order to better constrain these background components, we include an additional spectrum from the ROSAT All-Sky Survey from an annulus of $1-2^{\circ}$ around the center of A3667. This spectrum was extracted using the X-ray Background Tool\footnote{\url{heasarc.gsfc.nasa.gov/cgi-bin/Tools/xraybg/xraybg.pl}} provided by NASA's HEASARC. The normalizations for these components are left free. The spectral index for the CXB is fixed to $1.41$ \citep{2004A&A...419..837D}, while the temperatures for the LHB and GH were left free. For the absorbed components, the X-ray column density is fixed to $n_H = 4.31\times10^{-20}$~cm$^{-2}$, reported by the Leiden/Argentine/Bonn (LAB) Survey \citep{2005A&A...440..775K}. We also initially included a cool ($\sim 0.1$~keV) absorbed component to represent cooler Galactic emission \citep{2004ApJ...610.1182S}, but the normalization for this component was consistent with zero after fitting. For the column density reported here for this observation, this component is likely fully absorbed . The best-fit parameters for these components, fit simultaneously across all regions, are listed in Table~\ref{tab:BkgSpec}.

We include a number of additional instrumental background components. Two constants are included: one fixed to the solid angle for each region and detector, and another left free to account for small variations in the inter-detector calibration, linked across regions but left free across detectors. Gaussian emission lines that correspond to the Al K$\alpha$ and Si K$\alpha$ instrumental fluorescence lines are added for the MOS1 and MOS2 detector at $1.49$~keV and $1.75$~keV, and for the pn detector at $1.49$~keV. The line energies are fixed, the widths fixed to $0$, and the normalizations left free for each region and detector. 

We also include a separate model component for any residual soft proton emission (RESP) in the form of a power law that is not folded through the instrument effective area, following the ESAS cookbook. The spectral index and normalization for this component are left free but linked, with normalizations scaled appropriately, across all regions and detectors. The best-fit parameters for this component are listed in Table~\ref{tab:BkgSpec}.

An additional background component in the form of Solar Wind Charge Exchange (SWCX) is present in a small number of observations \citep{2004ApJ...610.1182S,2008A&A...489..837C,2011A&A...527A.115C}. To test if our observation is affected by SWCX, we use the method in \citet{2008A&A...489..837C}, where we examine the ratio of the low-energy ($0.5-0.7$~keV) light curve over the high-energy ($2.5-12$~keV) light curve. We find that the ratio is essentially flat over the entire observation. We therefore conclude that this observation is not significantly contaminated by SWCX and do not include additional model components to account for this. 

\begin{figure*}
\includegraphics[width=0.32\textwidth]{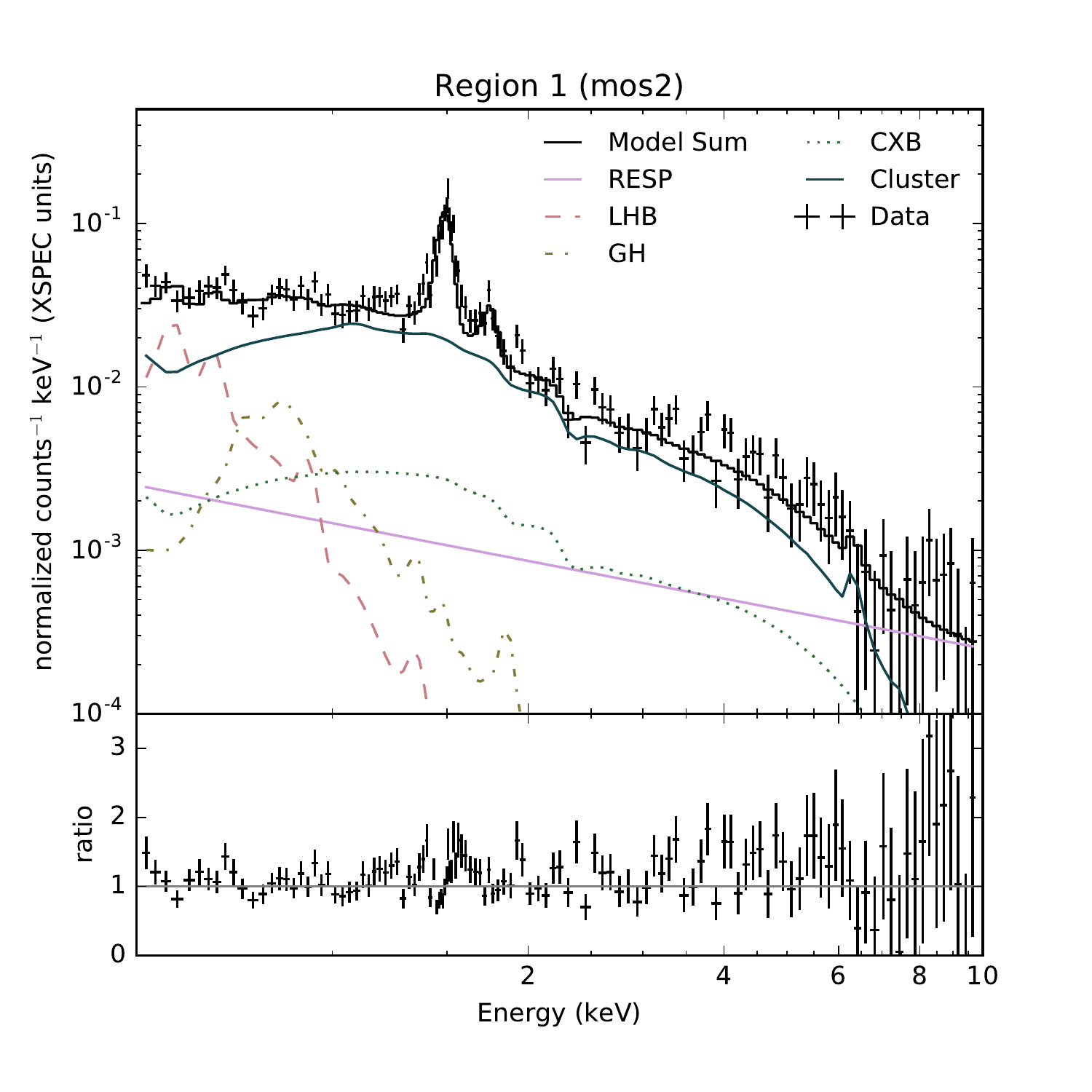}
\includegraphics[width=0.32\textwidth]{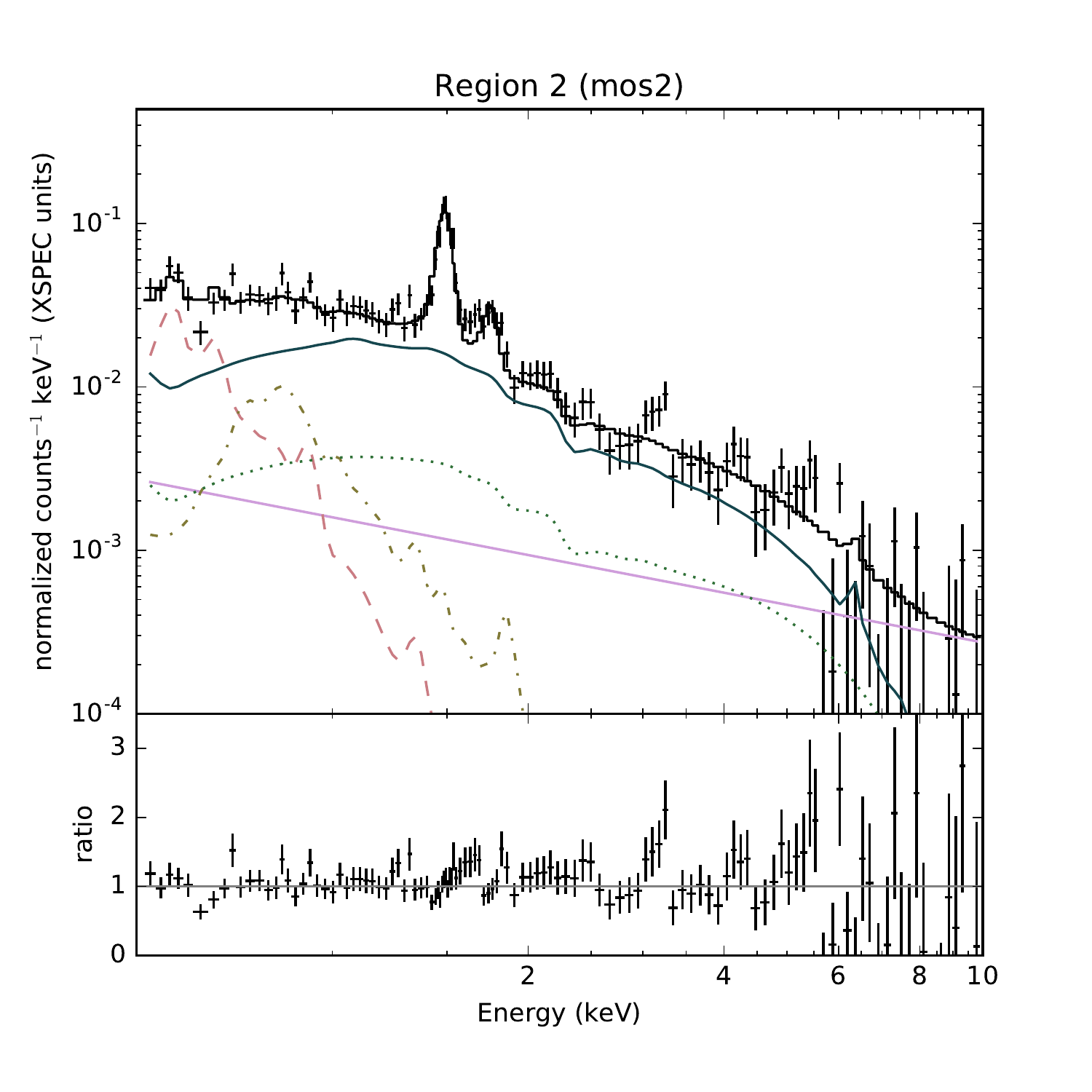}
\includegraphics[width=0.32\textwidth]{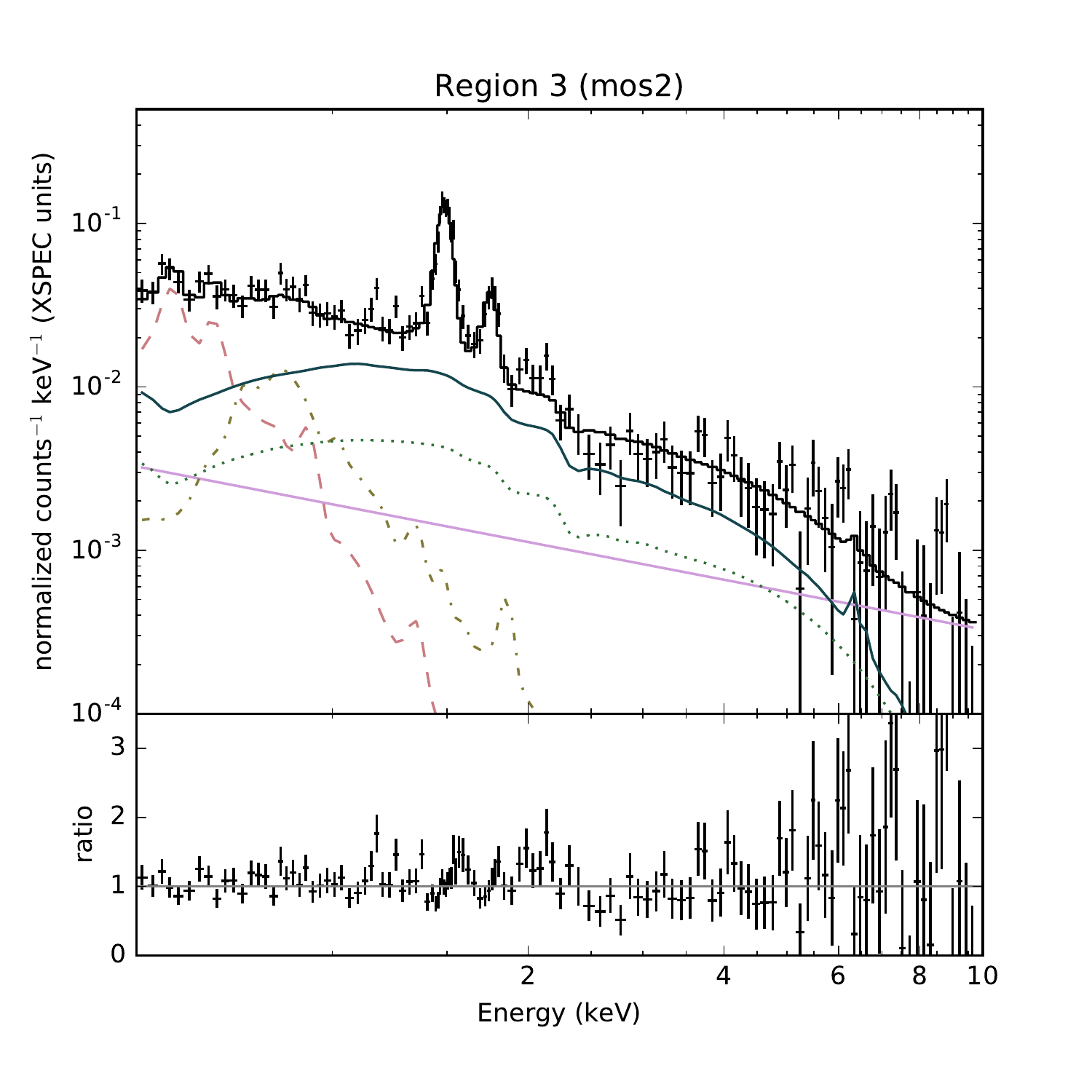}
\includegraphics[width=0.32\textwidth]{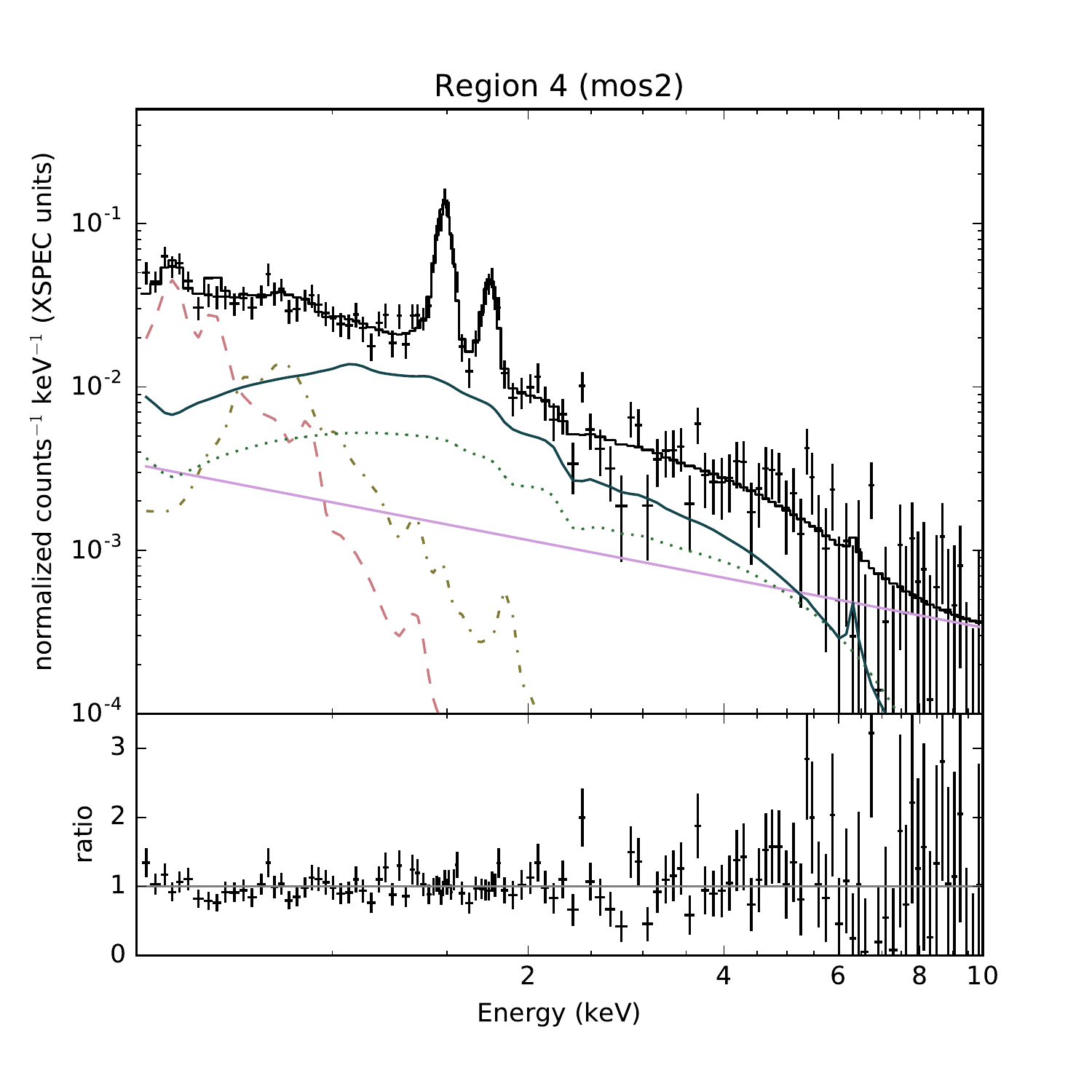}
\includegraphics[width=0.32\textwidth]{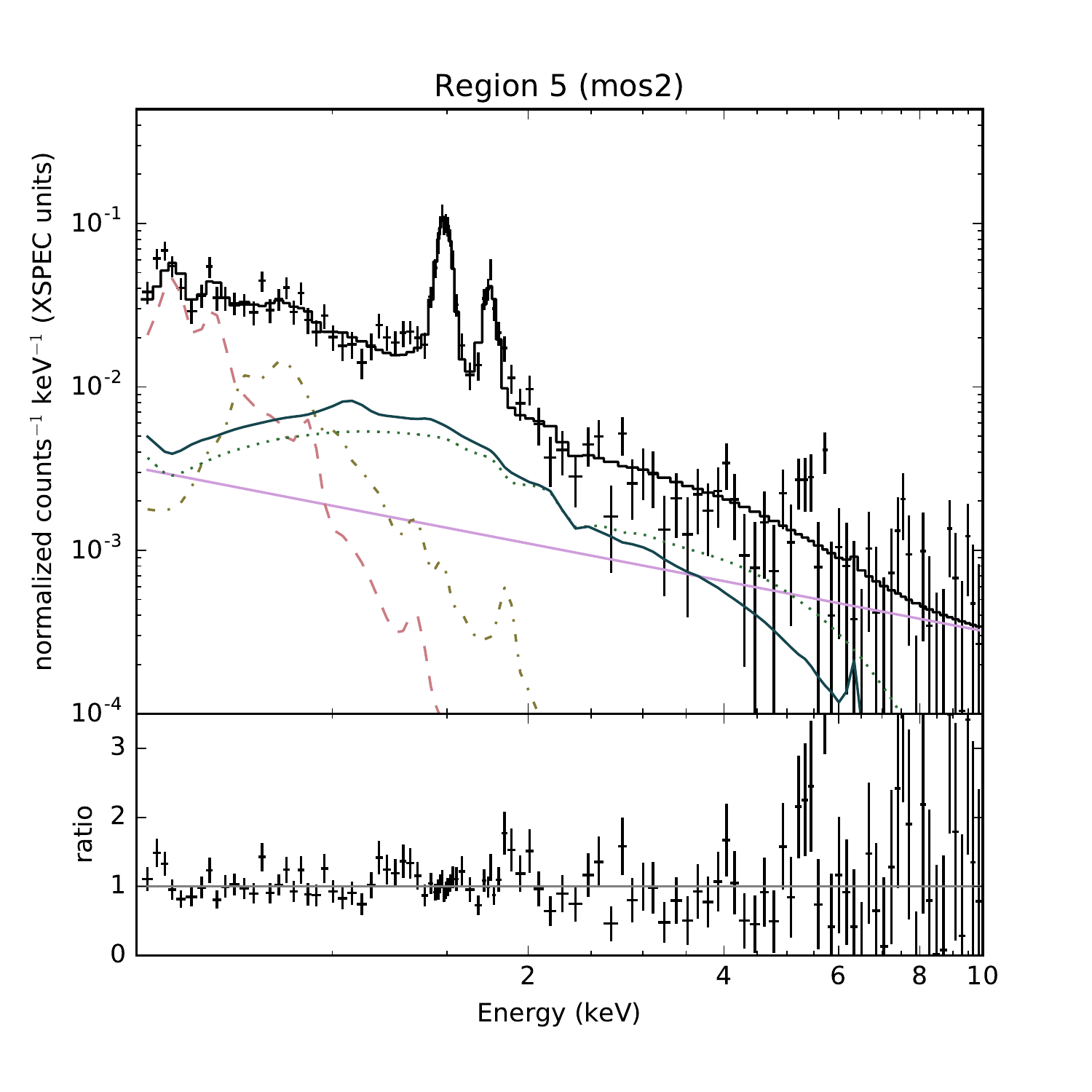}
\includegraphics[width=0.32\textwidth]{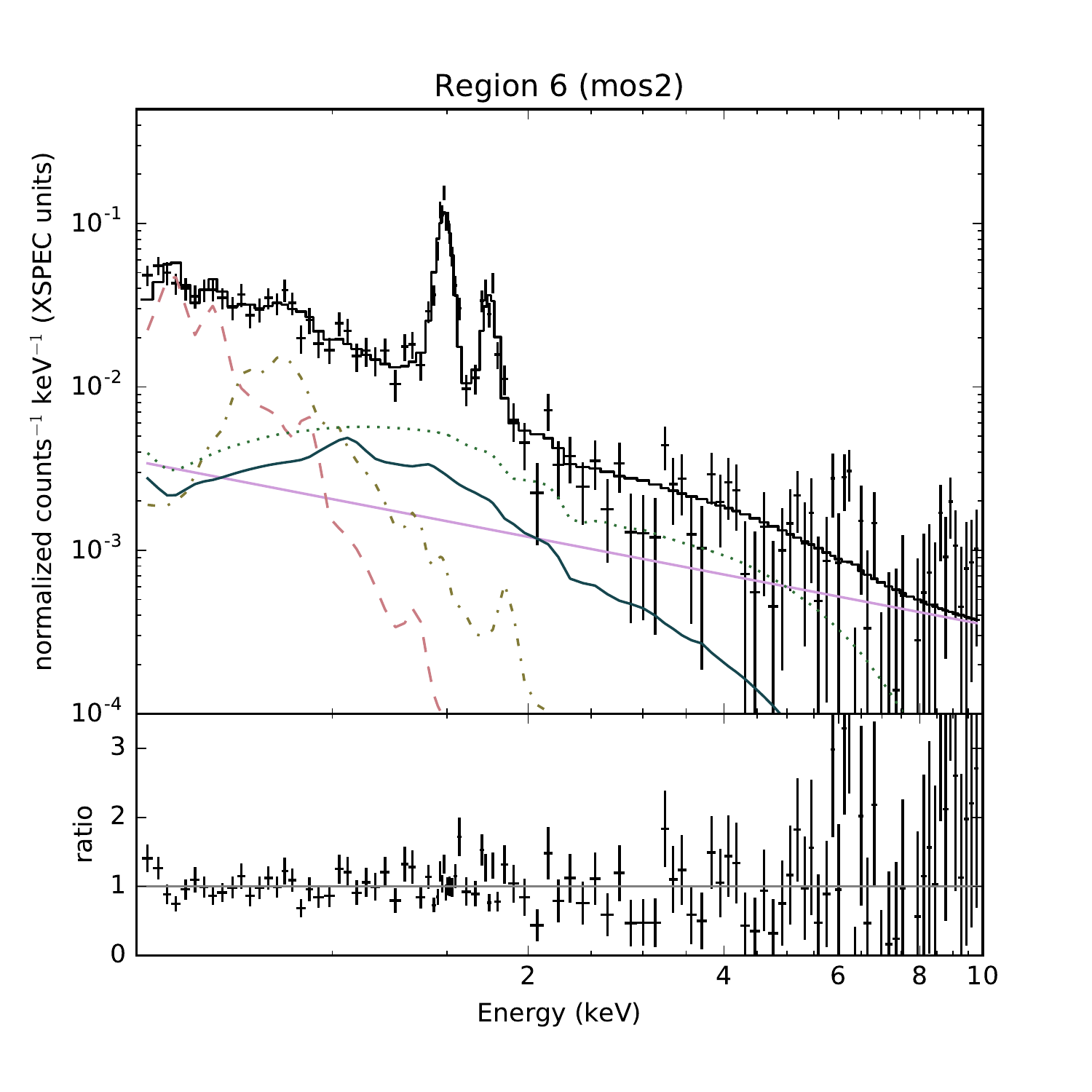}
\caption{Observed spectra with best-fit model components for the first six regions with nonzero cluster emission. For clarity, only the spectra for the MOS2 camera are shown; the spectra from the MOS1 and pn cameras look similar. Also for visual clarity, the black data points are grouped into bins with a minimum signal to noise ratio of 5. The plots are labeled in the same way as Figure~\ref{fig:SE_spectra_regions}. The solid black line is the model sum. The solid dark blue line is the cluster emission. The dotted green line is the CXB. The dashed orange and dash-dotted yellow lines are the LHB and GH, respectively. The solid pink line is the RESP. The RASS spectrum is not shown. The instrumental line fits at $1.49$~keV and $1.75$~keV are also not shown as separate model components, but are included in the overall fit.}
\label{fig:SE_spectra_grid}
\end{figure*}

\begin{table}
\centering
\caption{Cluster emission}\label{tab:ClusterSpec}
\begin{tabular}{lll}
	\hline
	Region & $kT$ & Normalization \\
    \hline
    1 & $5.6 \pm 0.5       $ & $(1.72 \pm 0.03) \times 10^{-5}$ \\
    2 & $5.8^{+0.6}_{-0.5} $ & $(1.14 \pm 0.21) \times 10^{-5}$ \\
    3 & $7.1^{+1.2}_{-0.9} $ & $(6.7 \pm 0.2) \times 10^{-6}$ \\ 
    4 & $5.0^{+0.8}_{-0.6} $ & $(5.5 \pm 0.2) \times 10^{-6}$ \\
    5 & $3.8^{+0.7}_{-0.5} $ & $(2.98 \pm 0.15) \times 10^{-6}$ \\
    6 & $2.7^{+0.8}_{-0.6} $ & $(1.49 \pm 0.15) \times 10^{-6}$ \\
    \hline
\end{tabular}
\begin{flushleft}
X-ray temperatures have units of keV. The normalizations are in XSPEC units, defined in Table~\ref{tab:BkgSpec}.
\end{flushleft}
\end{table}

\subsection{Results}
We fit the background and cluster emission spectra for all of the regions and detectors simultaneously. We are only able to extract temperatures from the first 6 regions, starting from the region closest to the cluster center (towards the NW). The normalizations of the cluster emission in the 4 remaining, SE-most regions are consistent with 0 at the $99\%$ confidence level. We therefore fix the cluster emission in these 4 regions to zero, but leave them in the fit to better constrain the background emission, especially any residual soft proton emission. We attempted to fit cluster emission to a larger region composed of the sum of these remaining 4 regions, but the normalization for this larger region was still consistent with 0 at the $99\%$ level.  

We find that there is some small but non-negligible remaining contribution from soft proton emission to the background. Leaving this component out results in a substantially worse fit ($\Delta \chi^2 = 293$ for a difference of 2 in the degrees of freedom). The fit is noticeably poorer especially for energies $\gtrsim6$~keV. 

We varied the cluster abundance from 0.1 to 0.5 (keeping this parameter fixed). There was essentially no change in the overall fit from 0.1-0.3 solar; higher values yielded slightly worse overall fits with little change in any of the parameter values. We also separately allowed the hydrogen column density to float. The preferred best fit value remained $n_H = 4.31\times10^{-20}$~cm$^{-2}$, consistent with the LAB measured value.

\begin{figure}
  \centering
  \includegraphics[clip,scale=0.7]{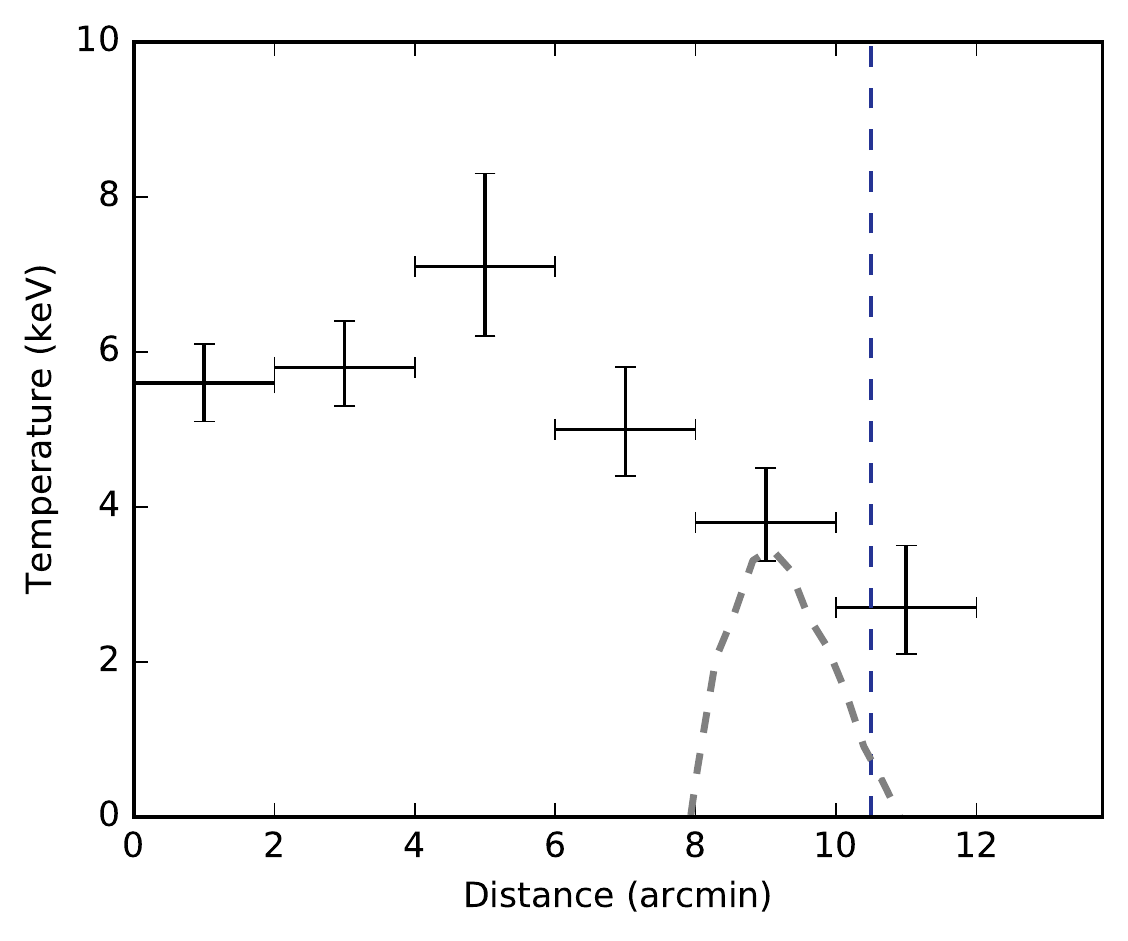}
  \caption{Temperature profile for regions $1-6$. Error bars on temperature are $1\sigma$ uncertainties; error bars on distance represent the width of each bin. The gray dotted line shows the SE radio relic surface brightness profile, in arbitrary surface brightness units and on a log scale (the peak is approximately 10 times brighter than the lowest value). The origin of the x-axis on this plot corresponds roughly to $r=5$~arcmin in Figure~\ref{fig:SE_SBprofile_wedge}. The vertical dashed blue line marks the approximate location of the best-fit shock radius from the spatial analysis.}
  \label{fig:TempProfile}
\end{figure}

We show the spectra and best-fit models for regions 1-6 in Figure~\ref{fig:SE_spectra_grid}. We plot the cluster temperature profile in Figure~\ref{fig:TempProfile}. The cluster temperatures and normalizations can be found in Table~\ref{tab:ClusterSpec}. This profile is roughly consistent with the temperature profile measured by \textit{Suzaku} by \citet{2013PASJ...65...16A}, which used different spectral regions and background modeling. 

There is a clear change in the slope of the temperature profile between regions 1-3 to 4-6. We can estimate the strength of the shock from the post- and pre-shock temperature change. Assuming a specific heat ratio of $\gamma=5/3$, the Rankine-Hugoniot jump conditions yield:

\begin{equation}
\frac{T_{post}}{T_{pre}} = \frac{5\mathcal{M}^4+14\mathcal{M}^2-3}{16\mathcal{M}^2}\,,
\end{equation}
where $\mathcal{M}$ is the Mach number, and $T_{post}$ and $T_{pre}$ are the post- and pre-shock temperatures, respectively. However, from Figure~\ref{fig:TempProfile}, it is not exactly clear where the edge of the shock front actually is. X-ray observations of relic regions in general suffer from projection effects, which may be responsible for smearing out the edge of a shock \citep{2013ApJ...765...21S,2015ApJ...812...49H}. If we take the trailing edge of the relic (region 4) to be post-shock with $T_{post} = 5.0^{+0.8}_{-0.6}$, and leading edge (region 6) to be pre-shock with $T_{pre} = 2.7^{+0.8}_{-0.6}$, the Mach number is then $1.8^{+0.5}_{-0.4}$. Within these statistical uncertainties, the estimate of the Mach number from the surface brightness is consistent with this estimate from the temperature jump (see however \citealt{2016arXiv160607433S} for a discussion of issues that can lead to discrepancies in Mach number estimates from surface brightness vs temperature profiles). We show this value along with estimates of the Mach number the surface brightness analysis and from other X-ray and radio observations, in Table~\ref{tab:MachNumber}. 

As seen in Figure~\ref{fig:SE_spectra_regions}, the shock front as inferred from the spatial analysis lies partially in regions 5 and 6. This provides further motivation for choosing to measure the Mach number between regions 4 and 6. This edge is also marked by a vertical, dashed blue line in Figure~\ref{fig:TempProfile} and lies at the leading edge of the radio relic (this line was measured by drawing a straight line from the center of the NW edge of spectral region 1 to the center of the arc in Figure ~\ref{fig:SE_spectra_regions}, which is located at the beginning of spectral region 6.)

If we instead take region 3 as the post-shock region, the Mach number is $2.4^{+0.6}_{-0.4}$, which is higher than but still consistent with our previous estimate as well as the estimate from \textit{Suzaku} and from DSA estimates. We therefore take our previous value, measured between regions 4 and 6, as a conservative estimate.

\begin{figure}
  \centering
  \includegraphics[clip,scale=0.25]{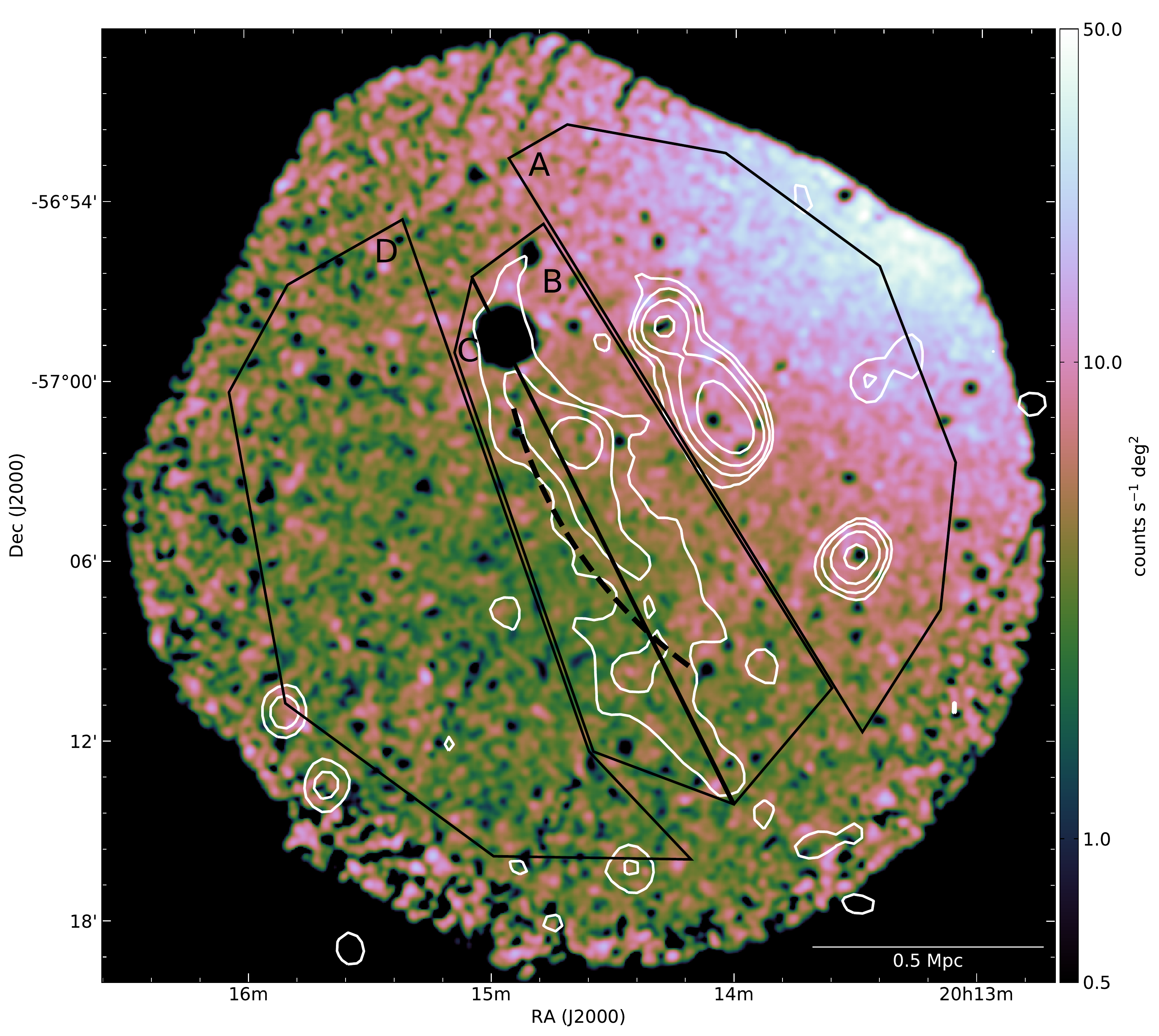}
  \caption{\textit{XMM} image of the SE region of A3667, $0.5-4$~keV, smoothed with a Gaussian kernel with a FWHM of $10$~arcsec. Overlaid in black are four regions used to cross-check the temperatures measured in the regions in Figure~\ref{fig:SE_spectra_regions}. The dotted black line marks the location of the best-fit shock radius from the spatial analysis. The white contours show the southern relic (with the same contour levels as in Figure~\ref{fig:XrayMosaic}).}
  \label{fig:SE_spectra_regions_2}
\end{figure}

As an independent check on this estimate, we also measure the temperatures in two regions that essentially split the relic in two non-overlapping halves lengthwise. We did not use the location of the density break to decide on these regions; we instead used only the shape of the radio emission to divide the relic region in half. As labeled in Figure~\ref{fig:SE_spectra_regions}, the post-shock region covers part of region 3 and the whole of region 4, and the pre-shock region covers all of region 6, and part of region 7, while region 5 is split between them. We show these regions in Figure~\ref{fig:SE_spectra_regions_2}. We find very similar temperatures to the ones listed in Table~\ref{tab:ClusterSpec}. The temperatures of the  post- and pre-shock regions (regions B and C) are $5.8^{+0.8}_{-0.7}$~keV and $3.2^{+1.0}_{-0.7}$~keV, respectively. The temperature in region A is $6.3 \pm 0.3$~keV. In region D, the normalization of the cluster emission is again consistent with 0 within $3\sigma$. The Mach number in this case, using the temperatures measured in regions B and C, is $1.8\pm0.4$. Therefore, we believe our temperature measurements and Mach number estimates are robust.

\section{Discussion}\label{sec:dis}
\begin{table*}
\centering
\caption{Mach number estimates for the SE relic of A3667.}\label{tab:MachNumber}
\begin{tabular}{llll}
\hline
	Mach Number & Method & Criteria & Reference \\
\hline
 	$2.45 \pm 0.26$ & Radio spectral index & $\alpha_{\mathrm{int}} = 0.9 \pm 0.1$ & \cite{2014MNRAS.445..330H} \\
	$3.3 \pm 1.5$ & Radio spectral index & $\alpha_{\mathrm{int}} = 1.2 \pm 0.2$, $\alpha_{\mathrm{inj}} = 0.7 \pm 0.2$ & \cite{2003PhDT.........3J} \\
	$1.75 \pm 0.13$ & X-ray temperature jump &($kT_{post}$, $kT_{pre}$) = ($6.34 \pm 0.38$,$3.59 \pm 0.28$) & \cite{2013PASJ...65...16A} \\
	$1.8^{+0.5}_{-0.4}$ & X-ray temperature jump &($kT_{post}$, $kT_{pre}$) =	$(5.0^{+0.8}_{-0.6}$, $2.7^{+0.8}_{-0.6}$) & this paper \\
	$1.3 \pm 0.1$ & X-ray surface brightness jump & $\mathcal{C}=1.4 \pm 0.2$ & this paper \\
    \hline
\end{tabular}
\begin{flushleft}
The second column `Method' indicates how the Mach number was obtained. The third column `Criteria' shows the measurements from observations used to calculate the Mach number. X-ray temperatures have units of keV. The radio spectral index is defined as follows: $S_{\nu} \propto \nu^{-\alpha}$, where $\alpha_{\mathrm{int}}$ is the integrated spectral index and $\alpha_{\mathrm{inj}}$ is the injected spectral index.
\end{flushleft}
\end{table*}

\subsection{Comparing estimates of the shock strength}

We summarize our estimations of the Mach number of the SE shock in A3667 in Table~\ref{tab:MachNumber}. Under the assumption of DSA, the Mach number is related to the radio spectral index as follows:
\begin{equation}
\mathcal{M} = \left(\frac{2\alpha_{\mathrm{inj}}+3}{2\alpha_{\mathrm{inj}}-1}\right)^{1/2}\,,
\end{equation}
where the radio injection spectral index, $\alpha_{\mathrm{inj}}$, is related to the electron injection spectral index, $\delta_{\mathrm{inj}}$, via $\alpha_{\mathrm{inj}} =(\delta_{\mathrm{inj}}-1)/2$, with $dN/dE = E^{-\delta_{\mathrm{inj}}}$ and $S_{\nu}\propto \nu^{-\alpha_{\mathrm{inj}}}$. In planar shocks, the radio injection index can be related to the volume integrated radio spectral index, $\alpha_{\mathrm{int}}$ via $\alpha_{\mathrm{int}} = \alpha_{\mathrm{inj}}+0.5$ \citep{1969ocr..book.....G}. However, this may not be a valid approximation for cluster shocks, which simulations suggest might be more spherical \citep[e.g.,][]{2015JKAS...48....9K,2015JKAS...48..155K}.

Ideally, the radio injection spectral index would be obtained directly from the detection of a gradient across the relic in a spatially-resolved spectral index map. However, this is often not possible due to the limited resolution in available observations. Observations of the SE relic with MWA by \citet{2014MNRAS.445..330H} did not display a spectral index gradient across the relic. Therefore, to calculate the Mach number, we will use the reported integrated index of $\alpha_{\mathit{int}}=0.9\pm0.1$ measured over the $120-226$~MHz range, with the acknowledgement that this is likely a lower limit on the Mach number. In this case the estimated Mach number is $\mathcal{M}=2.45\pm0.26$. An integrated spectral index of $\alpha_{\mathrm{int}}=1.2\pm0.2$ was measured at higher frequencies between $843$~MHz and $1400$~MHz, by \citet{2003PhDT.........3J}, where some evidence for a spectral gradient across the relic was found. A flatter spectral index of $\sim 0.5-0.7$ was measured at the outer edge of the radio relic. If we take the injected spectral index to be $\alpha_{\mathrm{inj}}=0.7\pm0.2$, the inferred Mach number would be $\mathcal{M}=3.3\pm1.5$. As is the case for several other relics with detected X-ray shocks, the shock strength estimated from DSA using the radio spectral index is higher than estimates from X-ray observations, but consistent with the X-ray measurements within 2$\sigma$. This discrepancy suggests that DSA is too simplistic to explain the nature of particle acceleration at cluster shocks. We discuss alternative scenarios in the next section, \ref{sec:dis2}.

A3667 is the second example of a double-relic cluster with deep X-ray observations that hosts one powerful relic and strong shock (the NW relic in this cluster) and one weaker relic and shock (the SE relic). These characteristics are also observed in CIZA J2242.8+5301 (the Sausage Cluster; \citealt{2015A&A...582A..87A}). An upcoming study indicates that a third cluster, A3376, displays this behavior as well (Urdampilleta, et al., in prep).\footnote{See also the conference presentation \citealt{2017xru..conf..229U}: \url{https://www.cosmos.esa.int/documents/332006/1402684/IUrdampilleta_t.pdf}} This behavior is likely driven by the specific dynamics of the cluster merger event. Simulations suggest that A3667 is post-merger \citep{1999ApJ...518..603R,2014ApJ...793...80D}. Optical observations support this and indicate that the merger axis is close to the plane of the sky \citep{2008A&A...479....1J}. Further analysis of the mass structure would help to confirm this scenario.

\subsection{Particle Acceleration at Radio Relics}\label{sec:dis2}

The acceleration of particles at shocks in astrophysical systems is typically attributed to DSA. However, it is generally assumed that DSA is not efficient for weak shocks $\mathcal{M}\lesssim 3$, because the injection into the DSA process is inefficient \citep{2012ApJ...756...97K}. Moreover, there appears to be a critical Mach number $\mathcal{M}=\sqrt{5}$ (corresponding to a compression ratio of $C=2.5$), below which cosmic-ray acceleration is not supported energetically \citep{2014ApJ...780..125V}. Indeed, shocks induced by coronal mass injections indicate that energetic particles are accelerated for all shocks above $C=2.5$ \citep{2012ApJ...761...28G}, whereas for lower compression ratios the situation appears to be more complicated. However, for the shocks in the solar system, the accelerated particles, though quite  energetic, are generally still sub-relativistic. In fact, the critical Mach number reported in  \citet{2014ApJ...780..125V} does depend on the adiabatic index of the accelerate particle population, with higher critical Mach numbers for fully relativistic particles.

A critical Mach number for the appearance of non-thermal radio emission from cluster shocks is quite appealing, as these shocks have Mach numbers $ 1< \mathcal{M} \lesssim 3$,  and, therefore, cap the range of the critical Mach number. However, our observations confirm that the SE relic in A3667 most likely has a Mach number below the critical Mach number, indicating that the  scenario of a critical Mach number controlling the presence or absence of non-thermal radio emission is too simple. The NW relic in A3367, for example, has a measured Mach number of $\sim 2.5$ \citep[e.g.,][]{2013PASJ...65...16A,2016arXiv160607433S}, above the critical threshold for DSA. However, while the X-ray emission is consistent with DSA as the main acceleration mechanism, the spectrum of the radio emission behind the X-ray shock is flatter than what DSA alone would predict \citep{2016arXiv160607433S}. Indeed, the majority of relics with confirmed X-ray shocks have Mach numbers that are $\lesssim 3$, with most of those $\lesssim 2.2$ \citep[e.g.,][]{2014MNRAS.440.3416O,2016ApJ...818..204V,2016MNRAS.461.1302E,2016MNRAS.460L..84B}.

The tension between DSA theory and the presence of radio relics associated with weak shocks has been recently addressed in the literature, with two possible scenarios for explaining the observations. One scenario is that the relativistic electrons are accelerated, perhaps even through second order Fermi acceleration, from a pre-accelerated population of electrons \citep{1998A&A...332..395E,2005ApJ...627..733M,2012ApJ...756...97K,2013MNRAS.435.1061P}, which helps to overcome the problem of injection of electrons into the DSA problem. Moreover, \citet{2014ApJ...780..125V} showed that pre-existing non-thermal particles (not necessarily electrons) helps to sustain the energetics of particle acceleration even below the critical Mach number. 

Evidence for the importance of reacceleration has recently been found for the merging clusters A3411/A3412b, where radio emission of the relic is associated with the passage of the shock through the trailing jet of a radio galaxy \citep{2017NatAs...1E...5V}. 

There are two bright radio sources located directly to the NW of the relic in A3667: PMN J2014-5701, a head-tail galaxy, and SUMSS J201330-570552, a radio source suspected to be a Farnoff-Riley type II galaxy \citep{2003PhDT.........3J}. PMN J2014-5701 was identified by \citet{2003PhDT.........3J} to have an optical counterpart that is likely a cluster member. If this is the case, this object could be a source of fossil electrons and support the hypothesis that relics associated with weak shocks are produced via reacceleration. However, there is no explicit connection or bridge between the emission from this source and the relic, which has been found in other examples of relics connected with fossil electron populations (e.g., the Bullet Cluster: \citealt{2015MNRAS.449.1486S}; A3411/A3412b: \citealt{2017NatAs...1E...5V}). A more detailed analysis of the radio emission, especially of a spectral index map, in combination with simulations of this cluster merger, could help to clarify this situation.

Even if secondary acceleration is indeed important, a question remaining then is if all relics need to be explained by reacceleration of pre-existing non-thermal electrons, or whether relics are produced by a mixture of DSA and reacceleration, with the weakest shocks, such as the one associated with the A3667 SE relic, producing non-thermal electrons solely caused by reacceleration.

Another scenario to explain the presence of non-thermal particles associated with weak cluster shocks is to take into account the possible variations in Mach number along the shock \citep[e.g.,][]{2012MNRAS.421.1868V,2013ApJ...765...21S,2015ApJ...812...49H,2017arXiv170707085K}. Relics can be $\sim$Mpc in size, and it is unlikely that the average shock Mach number inferred from X-ray observations exactly holds at each position along the shock: the shock velocity may vary along the shock, but also the pre-shock temperature will show some variation. As  a result there may be parts of the shock structure where the Mach number is above the critical Mach number and particle acceleration is efficient. 

Based on the current X-ray and radio observations it is difficult to test this latter scenario, but  ESA's future Athena X-ray mission \citep{2013arXiv1306.2307N} may be able to better measure the variation in upstream temperature to probe the possible variation in Mach numbers. 

Last but not least, an additional complication with invoking DSA in general is the lack of observed gamma-ray emission from cosmic ray protons which should also be accelerated at shocks \citep[e.g.,][]{2014ApJ...787...18A,2016ApJ...819..149A,2016A&A...589A..33A,2016MNRAS.459...70V}. In particular, applying standard DSA to explain radio relics should yield a total cluster-wide gamma-ray emission at a level that should have already been detected by \textit{Fermi} \citep{2014MNRAS.437.2291V}, or else require unrealistically large magnetic fields and/or cosmic ray proton-to-electron number ratios \citep{2015MNRAS.451.2198V}. Particle-in-cell simulations indicate that protons and electrons are accelerated differently depending on the shock geometry \citep{2014ApJ...794..153G,2014ApJ...797...47G,2014ApJ...783...91C}. When taken into account in simulations of relics in clusters, this effect can alleviate the tension between the expected gamma-ray emission and observed upper limits in certain cases \citep{2017MNRAS.464.4448W}. The already-mentioned possible presence of a fossil population of cosmic ray electrons could also alleviate this problem \citep{2014MNRAS.437.2291V}.

\section{Conclusion}\label{sec:con}

We present new observations of the SE relic region of the merging cluster A3667. We find a clear temperature jump across the radio relic, consistent with previous observations from \textit{Suzaku}. We find weak evidence for a corresponding surface brightness discontinuity. We interpret these discontinuities, along with the presence of the radio relic, as evidence for a shock, and estimate a shock strength of $\mathcal{M}\approx 1.8$, consistent with previous X-ray and radio estimates. This cluster is another example of a system with one powerful radio relic and strong X-ray shock and one weak relic and shock.

This low Mach number is not consistent with the idea that for diffusive shock acceleration to accelerate particles the Mach number has to be $M\gtrsim 2.2$. Possible solutions to reconcile DSA with the presence of non-thermal electrons at a low Mach number shocks are reacceleration of pre-existing accelerated electrons, or to allow for excursions of the  Mach number along the shock, which on average may be  $\mathcal{M}\approx 1.8$, but with isolated regions with  $M\gtrsim 2.2$. The future X-ray mission Athena may be able to provide clarity to this issue.

\section*{Acknowledgements}

E.S. acknowledges support from NWO through a Vidi grant (PI: C. Weniger). H.A. and F.Z. acknowledge the support of NWO via Veni grants.





\bibliographystyle{mnras}
\bibliography{references} 

\bsp	
\label{lastpage}
\end{document}